\documentclass[runningheads]{llncs}
\usepackage[T1]{fontenc}
\usepackage{lmodern}
\usepackage{graphicx}
\usepackage{ulem}
\usepackage{verbatim}
\usepackage{multirow}
\usepackage[lined,algonl,boxed]{algorithm2e}
\usepackage{longtable}
\usepackage{array}
\usepackage{subfigure}
\usepackage{stfloats}
\usepackage{balance}
\usepackage{multicol}
\usepackage{color}
\usepackage{bm}
\usepackage{booktabs} 
\usepackage{epsfig}
\usepackage{enumitem}

\usepackage{arydshln}

\usepackage{amsmath,amssymb,amsfonts}
\usepackage{cleveref}
\usepackage{algorithmic}
\usepackage{graphicx}
\usepackage{textcomp}
\usepackage{xcolor}
\usepackage{subfigure}
\usepackage[misc]{ifsym}

\usepackage{tabularx}

\newcommand{\hide}[1]{} 
\newcommand{\vpara}[1]{\vspace{0.1in}\noindent\textbf{#1}}
\newcommand{\ipara}[1]{\vspace{0.1in}\noindent\textit{#1 }}

\newcommand{\beq}[1]{\vspace{-0.02in}\begin{equation}#1\end{equation}\vspace{-0.02in}}

\begin{document}

\title{Recommending Courses in MOOCs for Jobs: An Auto Weak Supervision Approach}

\titlerunning{AutoWeakS: Recommending Courses in MOOCs for Jobs}
\authorrunning{B. Hao et al.}
%
\author{Bowen Hao\inst{1}\and
	Jing Zhang\inst{1}\and
	Cuiping Li\inst{1} \and
	Hong Chen\inst{1}\and
	Hongzhi Yin\inst{2}}
%
%
\institute{Key Laboratory of Data Engineering and Knowledge Engineering of Ministry of Education, School of Information, Renmin University of China\and
	School of Information Technology and Electrical Engineering, The University of Queensland \\
	\email{\{jeremyhao, zhang-jing, licuiping, chong\}@ruc.edu.cn\\ h.yin1@uq.edu.au\\}
}

\maketitle            
\begin{abstract}
The proliferation of massive open online courses (MOOCs) demands an effective way of course recommendation for jobs posted in recruitment websites, especially for the people who take MOOCs to find new jobs. Despite the advances of supervised ranking models, the lack of enough supervised signals prevents us from directly learning a supervised ranking model. This paper proposes a general automated weak supervision framework (\textit{AutoWeakS}) via reinforcement learning to solve the problem. On the one hand, the framework enables training multiple supervised  ranking models upon the pseudo labels produced by multiple unsupervised ranking models. On the other hand, the framework enables automatically searching the optimal combination of these supervised and unsupervised models. \quad Systematically, we evaluate the proposed model on several datasets of jobs from different recruitment websites and courses from a MOOCs platform. Experiments show that our model significantly outperforms the classical unsupervised, supervised and weak supervision baselines.

\hide{  }
\end{abstract}

\section{Introduction}

Massive open online courses, or MOOCs, are attracting widespread interest as an alternative education model. Lots of MOOCs platforms such as Coursera, edX and Udacity have been built and provide low cost opportunities for anyone to access a massive number of courses from the worldwide top universities. As reported by Harvard business review\footnote{ 
		 {\scriptsize https://hbr.org/2015/09/whos-benefiting-from-moocs-and-why}}, a primary goal of 52\% of the  people surveyed who takes MOOCs is to improve their current jobs or find new jobs. We call this group of MOOCs' users as career builders. Meanwhile, people usually  resort to the online recruitment platforms such as LinkedIn.com and Job.com to seek jobs. However, there always exists a ``Skill Gap"~\cite{Tong2017Measuring} between the career builders and the employers. The career builders who expect themselves to fit a job through taking MOOCs, need to deeply understand the demands of the job skills and then take the matchable courses. Clearly, to help career builders improve their skills for finding gainful jobs, it has been an essential task that is able to automatically match jobs with suitable courses.

Straightforwardly, to solve this problem, unsupervised methods such as BM25~\cite{robertson2009probabilistic}, Word2vec~\cite{mikolov2013distributed} or  the network embedding methods such as DeepWalk~\cite{perozzi2014deepwalk} and LINE~\cite{line-large-scale-information-network-embedding} can be used to calculate the relevance between a queried job and a candidate course. However, such unsupervised methods aim at  modeling the implicit structures of the input data, i.e., the clustering of words in jobs and courses, while the ranking of different courses to a queried job cannot be obviously learned.  In another word, the unsupervised models do not explicitly compare the relevance of the positive courses and the negative courses  to a queried job. Although the supervised neural ranking models are demonstrated to have good performance  in the information retrieval (IR) tasks~\cite{qiu2015convolutional,pang2016text}, they cannot  directly solve our problem, as the supervision signals about which courses can be recommended to a job are not easily available.

To alleviate the problem of lacking supervision signals, weak supervision models are proposed to  train supervised IR models upon the pseudo labels provided by unsupervised models. For example, Dehghani et al. leverage the output of BM25 as the weak supervision signals~\cite{dehghani2017neural} and Zamani and Croft extend a single pseudo signal to multiple signals to guide multiple supervised ranking models~\cite{zamani2018neural}. However, for different tasks, human efforts are demanded to determine the suitable weak signals and the supervised models. Even if each component is carefully selected by humans, their combination may not result in the best performance (which is also justified in our experiments). 
Thus, it is imperative to automatically identify an optimal combination of different components. 

To address the above challenge, we propose a general automated  weak supervision model \textit{AutoWeakS}, which can automatically select the optimal combination of weak signals, supervised models and hyperparameters for a given ranking task and dataset. Specifically, the auto model trains a weak supervision model and a controller iteratively through reinforcement learning, where the weak supervision model aims to train a group of sampled supervised ranking models upon the pseudo labels (i.e., weak signals) provided by a group of sampled unsupervised ranking models, and the controller targets at automatically sampling an optimal configuration for the weak supervision model, i.e., it sequentially determines which unsupervised models should be sampled, how to set the hyperparameters for merging the unsupervised models, and which supervised models should be sampled. Our proposed model is a general framework to rank courses for jobs in this paper, but it is general enough to solve other ranking problems. Besides, we can incorporate any unsupervised and supervised models as candidate components to be selected by the controller. 

Our contributions can be summarized as: (1) we are the first to explore the problem of recommending courses in MOOCs to jobs posted in online recruitment websites, which can help to eliminate the ``Skill Gap" between the career builders who take MOOCs and the employers in the recruitment; (2) we propose a general automated weak supervision model, \textit{AutoWeakS}, to rank courses for jobs. With reinforcement joint training of the weak supervision model and the controller in \textit{AutoWeakS}, we can automatically find the best configuration of the weak supervision model; (3) experiments on two real-world datasets of jobs and courses show that \textit{AutoWeakS} significantly outperforms the classical unsupervised, supervised and weak supervision baselines.


\begin{figure*}[t]
	\centering
	\includegraphics[width=1.0\textwidth]{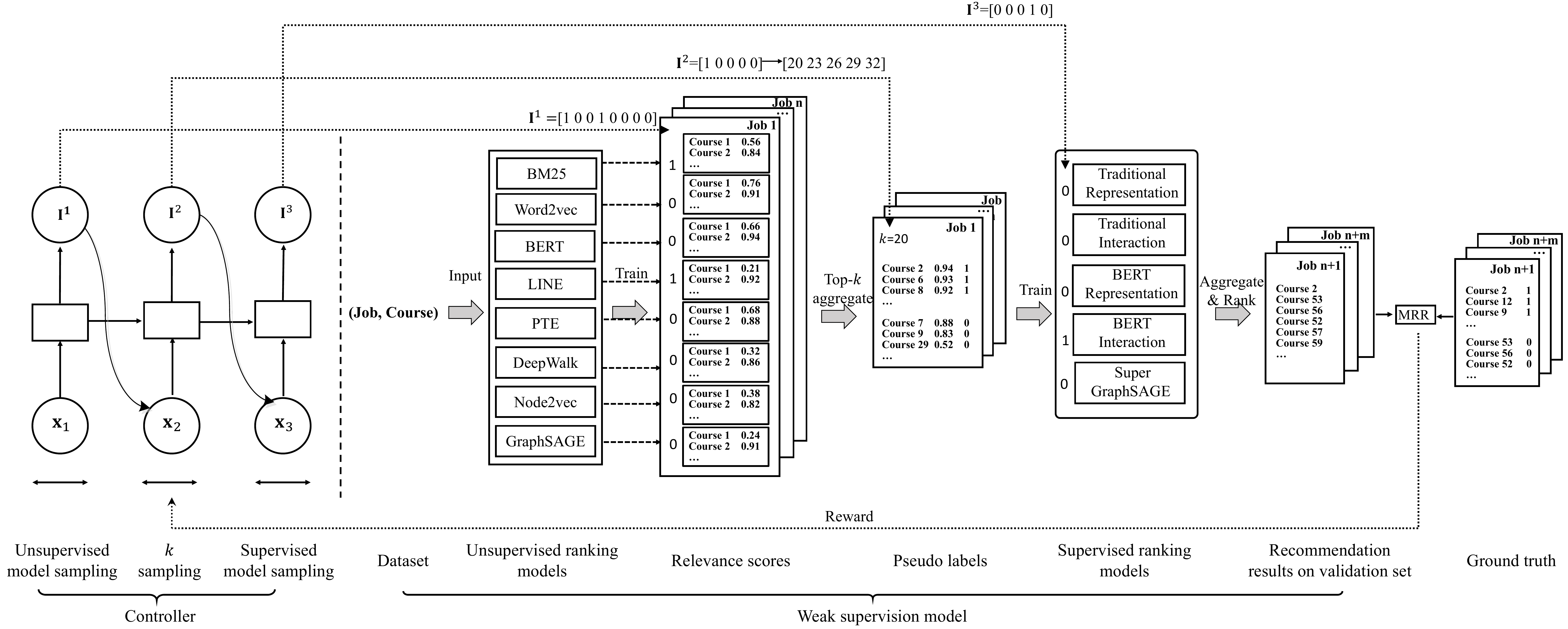}
	\caption{\label{fig:overview} Overview of the auto weak supervision model.}
\end{figure*}

\section{The Auto Weak Supervision Model}
\label{proposed_model}

We denote the jobs in a recruitment website as $J$ and the courses in a MOOCs platform as $C$, where each job $j \in J$ contains maximal $N$ words, i.e., $j = \{j_{1}, \cdots, j_{N}\}$, and each course $c \in C$ contains maximal $M$ words, i.e., $c= \{c_{1}, \cdots,  c_{M}\}$. The words are extracted  from the descriptions of the jobs or courses. Given $J$ and $C$, the goal is to learn a predictive function $\mathcal{F}: (J, C) \rightarrow Y$ to predict the label $y \in Y$ for each pair of a queried job $j$ and a candidate course $c$, where $y$ is a binary value with $y =1$ if $c$ is relevant to $j$ and $y=0$ otherwise. 

\subsection{Model Overview}
In our problem, the set of the true labels $Y$ about which courses should be recommended to a given job are not easily to be obtained, which motivates us to use a weak supervision method to solve this task, i.e., training supervised ranking models upon the pseudo labels provided by unsupervised models. 
However, selecting only one unsupervised model may suffer from the issue of ranking bias, while combing multiple unsupervised models may bring in additional noises. Besides, we also have many different choices for the supervised ranking models. Thus we explore an optimal combination of the pseudo labels from various unsupervised ranking models, together with various supervised ranking models.

Fig.~\ref{fig:overview} illustrates the proposed auto weak supervision framework \textit{AutoWeakS}, which consists of a weak supervision model and a controller, where the weak supervision model aggregates the results of multiple unsupervised ranking models as the pseudo labels and trains multiple supervised ranking models upon them, and the controller  is responsible for automatically searching the optimal configuration of the unsupervised and the supervised models. Specifically, first, the controller is to sample the unsupervised models, sample the number $k$ of the top ranked courses to a job after aggregating the results of the unsupervised models, and sample the supervised models to be trained upon the top-$k$ pseudo labels, and since the above three sampling processes should be sequentially determined, we formalize the controller by a three-step LSTM model; 
second, the sampled supervised models are trained on the pseudo labels and evaluated on the validation data with a few human annotated labels; finally, the evaluation metric is returned as the reward signal to guide the training of the controller. When the whole training process converges, we can obtain an optimal combination of different components, which can be regarded as the final model to predict the courses for new jobs. 

\subsection{Weak Supervision Model}
\label{weak-supervision}

In the weak supervision model, we conduct $N^u$ unsupervised ranking models to calculate $N^u$ relevance scores for each pair of a queried job and a candidate course, aggregate the $N^u$ relevance scores to generate the pseudo labels, and train $N^s$ supervised ranking models on these pseudo labels. Although we select the following unsupervised and supervised models in our framework, the framework is general to incorporate any kinds of unsupervised and supervised models.

\vpara{Unsupervised Ranking Model.}
\label{sub:unsupervised-model} We define two types of the unsupervised ranking models, namely unsupervised text-only matching models and unsupervised graph-based matching models. Unsupervised text-only matching models calculate a relevance score between a queried job and a candidate course based on their descriptions. For example, \textit{BM25}~\cite{robertson2009probabilistic}  exactly matches the words between a job and a course. \textit{Word2vec}~\cite{mikolov2013distributed} and \textit{BERT}~\cite{devlin2018bert} first embed the descriptions of a course and a job into two vectors, and then calculate the cosine similarity between these two vectors.

Different from the unsupervised text-only matching models  which represent the jobs and the courses independently, unsupervised graph-based matching models leverage the global correlations between the jobs and courses to represent them. Specifically, we first build a job-word-course heterogeneous graph $G=(V,E)$, which consists of three types of nodes, i.e., job, course and word, and two types of edges that connect courses and words, and connect jobs and words. Then we apply different unsupervised network embedding models to map each node in $G$ into a low-dimensional vector to capture the structural properties. For example, \textit{LINE}~\cite{line-large-scale-information-network-embedding} and PTE ~\cite{Jian2015PTE} maximize the first-order and the second-order proximity between two nodes. 
\textit{DeepWalk}~\cite{perozzi2014deepwalk} extends the first-order neighbors to distinct neighbors which can be reached by random walks. \textit{Node2vec}~\cite{Grover2016node2vec} further proposes the biased random walks to balance the homophily by BFS search and the structural equivalence by DFS search. 
\textit{GraphSAGE}~\cite{hamilton2017inductive} aggregates the neighbors' embeddings of the nodes to represent them.
Finally, we can calculate the relevance based on the learned embeddings of job and course.

\vpara{Pseudo Label Generator.}
To avoid the labeling bias from a single unsupervised model, we aggregate the results of the $N^u$ unsupervised ranking models to generate the pseudo labels. Specifically, for each queried job, we average the $N^u$ relevance scores for each candidate course, rank all the courses according to their average relevance scores,  and then annotate the top-$k$ courses as positive instances and the other courses as the negative instances for the queried job.

\vpara{Supervised Ranking Model.}
We define two types of the supervised ranking models, including supervised  text-only matching models and supervised graph-based matching models. For the supervised text-only matching models, we first explore two traditional models, namely traditional representation model and  traditional interaction model, and inspired by the recently proposed pre-training model BERT~\cite{devlin2018bert}, which has advanced the state-of-the-art in various NLP tasks, we further explore two BERT-based models, namely BERT representation model and BERT interaction model. Finally, we explore one supervised graph-based matching model,  GraphSAGE.

\ipara{Traditional Representation Model} directly compares the embeddings of a queried job and a candidate course to capture their semantic relevance. Fig~\ref{subfig:representation} illustrates the architecture of the model. We first transform the input word representations $\mathbf{x}  \in \mathbb{R}^{N \times d_0}$ of a job  into low-dimensional embeddings, and then apply multi-layer nonlinear projections on them to get the intermediate embeddings $\mathbf{h}_l \in \mathbb{R}^{d_l}$ and the final embedding $\mathbf{y} \in \mathbb{R}^{d_L}$  of a job, where $N$ is the maximal number of words included in all the jobs, $d_0$, $d_l$ and $d_L$ represent the embedding dimensions. The paired inputs of courses are transformed in the same way. Finally, we estimate the relevance score $r(j,c)$ of $c$ to $j$ as the cosine similarity between the job embedding $\mathbf{y}_{j}$ and the course embedding $\mathbf{y}_c$. 

\begin{figure}[t]
	\centering
	\mbox{ 
		\subfigure[\scriptsize Traditional Representation Model]{\label{subfig:representation.pdf}
			\includegraphics[width=0.405\textwidth]{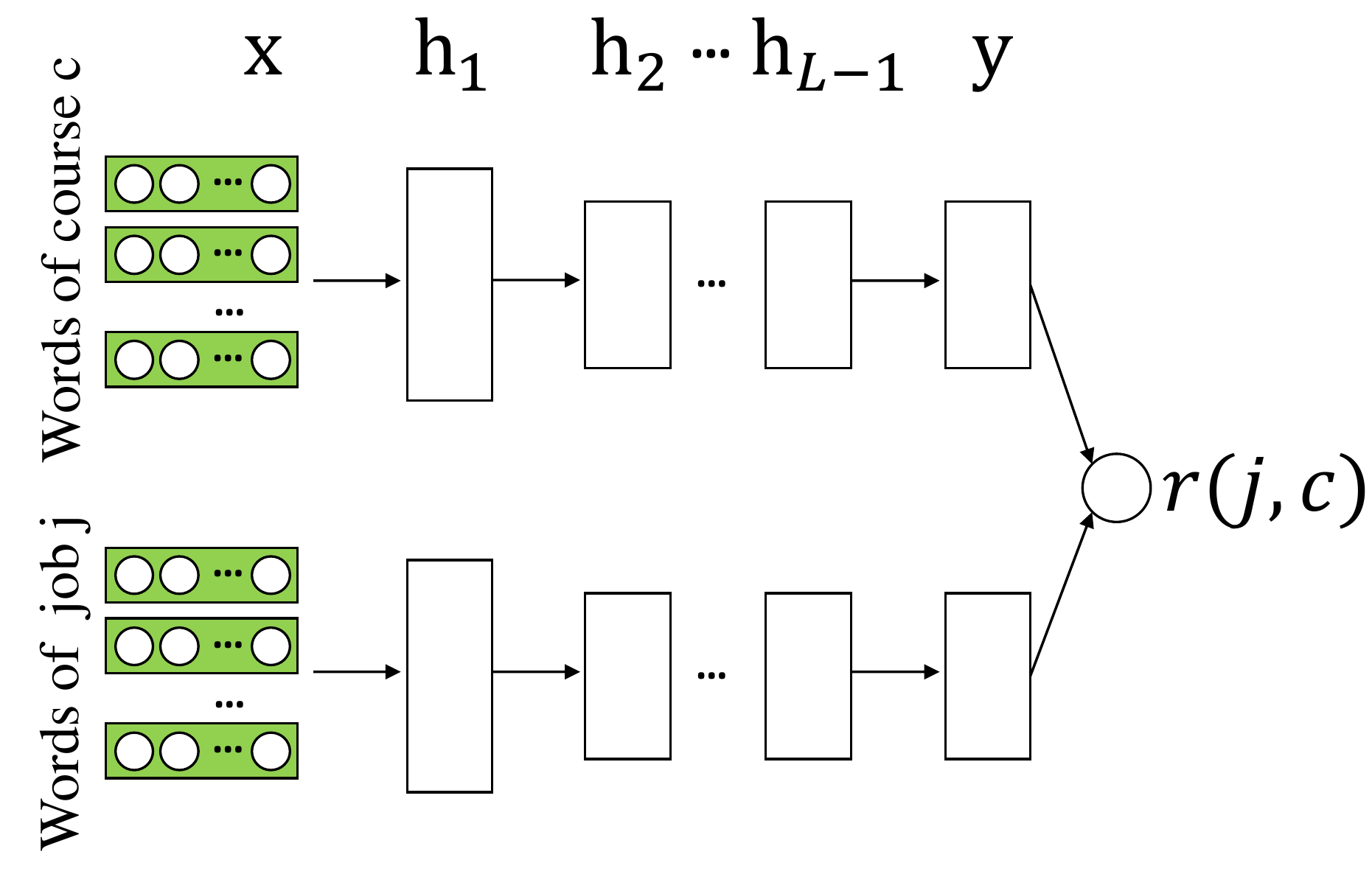}
		}
		
		\hspace{+0.15in}
		
		\subfigure[\scriptsize Traditional Interaction Model]{\label{subfig:interaction.pdf}
			\includegraphics[width=0.405\textwidth]{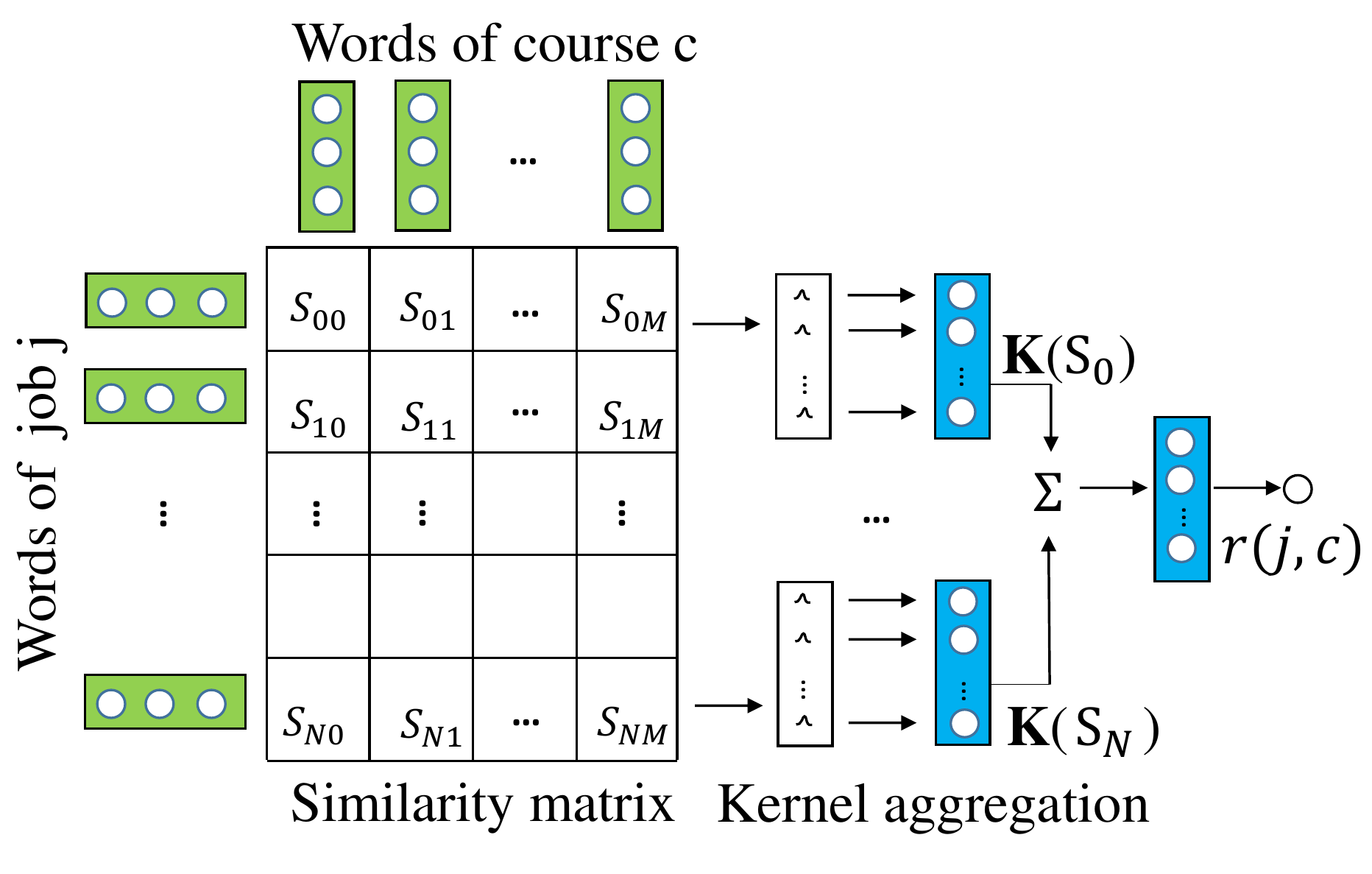}
		}

	}

	\caption{\label{fig:neural_ranking_model} Traditional supervised text-only matching models.  }
\end{figure}

\begin{figure}[t]
	\centering
	\mbox{ 
		\subfigure[\scriptsize BERT Representation Model]{\label{subfig:bert_representation.pdf}
			\includegraphics[width=0.405\textwidth]{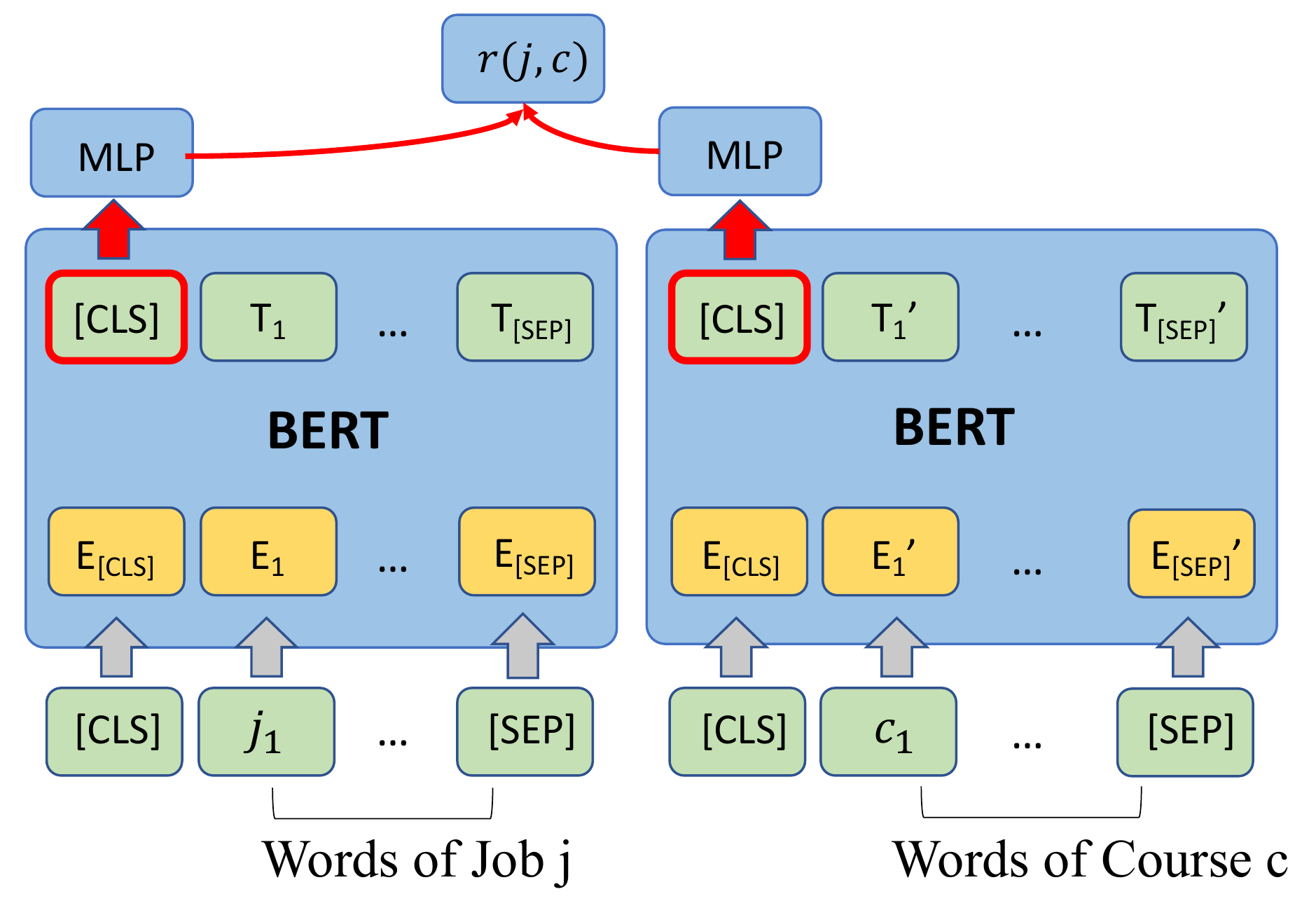}
		}

		\subfigure[\scriptsize BERT Interaction Model]{\label{subfig:bert_interaction.pdf}
			\includegraphics[width=0.405\textwidth]{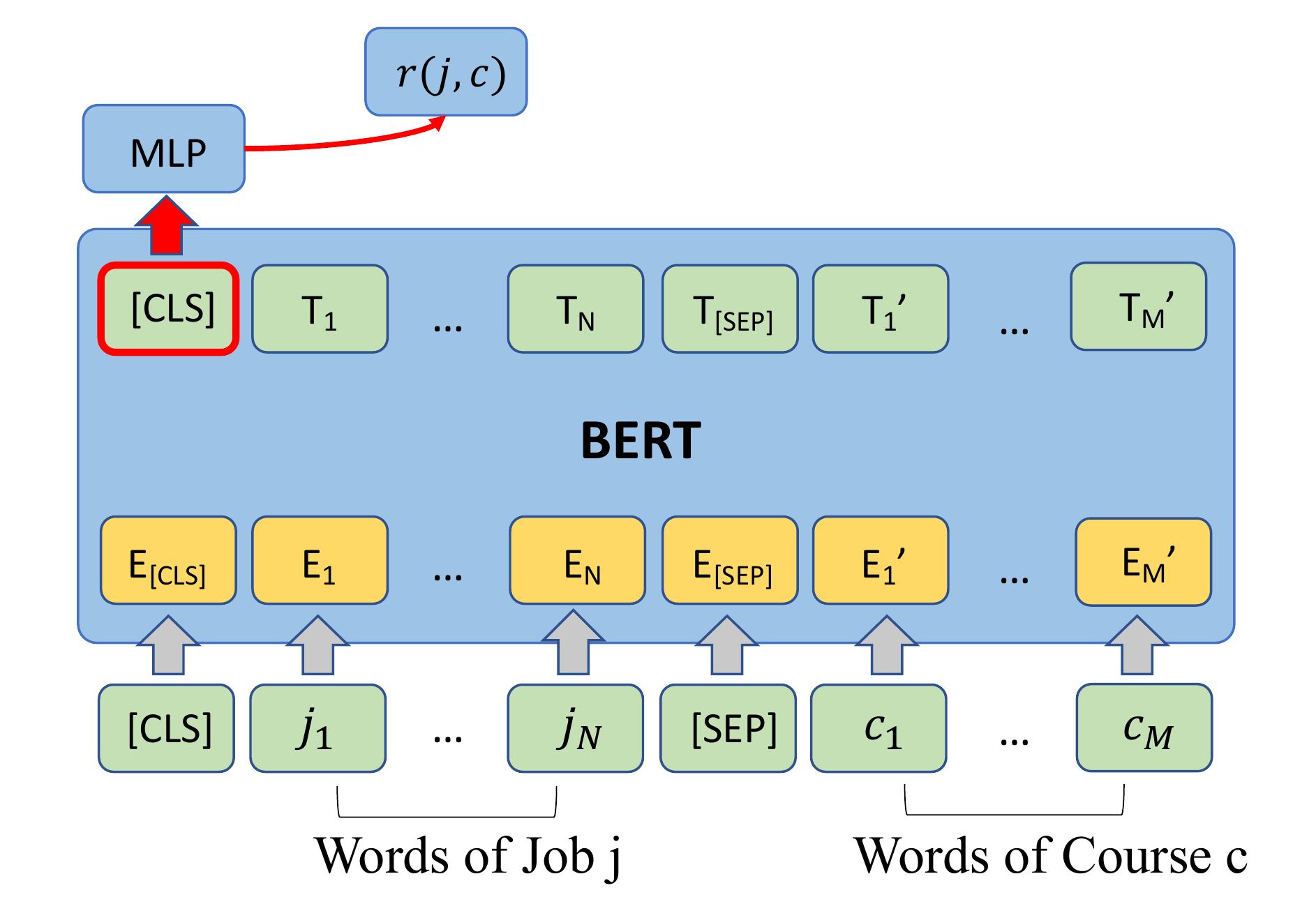}
		}

	}

	\caption{\label{fig:neural_ranking_mode1l} BERT-based supervised text-only matching models.  }
\end{figure}

\ipara{Traditional Interaction Model} compares each pair of the words in a queried job and a candidate course. Fig.~\ref{subfig:interaction} illustrates the model. Inspired by~\cite{xiong2017end}, we first build a similarity matrix $\mathbf{S}$ between the word embeddings of a queried job and a candidate course, where each element ${S}_{ik}$ in the similarity matrix $\mathbf{S}$ stands for the cosine similarity between the embedding of the $i$-th word in job $j$ and the embedding of the $k$-th word in course $c$. Then we transform the $i$-th row $\mathbf{S}_i = \{{S}_{i0}, \dots,  {S}_{iM}\}$ of the similarity matrix $\mathbf{S}$ into a feature vector $\mathbf{K}(\mathbf{S}_i) = \{{K}_1(\mathbf{S}_i), \dots, {K}_H(\mathbf{S}_i)\}$, where $\mathbf{S}_i = \{{S}_{i0}, \dots,  {S}_{iM}\}$ represents the similarities between the $i$-th word of the queried job $j$ and every word of the course $c$. Each of the $h$-th element is converted from $\mathbf{S}_i$  by the $h$-th RBF kernel with the mean value $\mu_h$ and the variance value $\sigma_h$, i.e., ${K}_h(\mathbf{S}_i)= \sum_{k=1}^M \exp[({S}_{ik}-\mu_h)^2/2\sigma_h^2]$. 
Next, the similarity vectors of all the words in $j$ are summed up into a similarity feature vector, i.e.,  $\sum_{i=1}^N{\log}\mathbf{K}(\mathbf{S}_i)$, which is then mapped into a one-dimension relevance score $r(j,c)$ to represent the relevance between the job $j$ and the course $c$.  

\ipara{BERT Representation Model} compares the embeddings of a queried job and a candidate course through independently encoding the descriptions of a queried job and a candidate course by the pre-training model BERT~\cite{devlin2018bert}. Fig.~\ref{subfig:bert_representation} illustrates the proposed model. Specifically, for each job $j$, we take [CLS], $j_1,\cdots,j_N$, [SEP] as the input, where $j_1,\cdots,j_N$ represent job tokens, [CLS] and [SEP] are special tokens. Then we add a multi-layer perceptron (MLP) layer on top of the first output [CLS] embedding to get the representation of job $j$. A course $c$ is represented in the same way. Finally, we calculate the cosine similarity between the embeddings of the job and the course to obtain the relevance score $r(j,c)$.

\ipara{BERT Interaction Model} compares each pair of the words in a queried job and a candidate course through a unified BERT model, where the multi-head attention in the BERT unit spans over the interactions between the job and the course so that the job-course interactions  can be captured. Fig.~\ref{subfig:bert_interaction} illustrates the proposed model.  Specifically, we take [CLS], $j_1, \cdots, j_N,$ [SEP], $c_1,\cdots,c_M$ as the input, where $j_1,\cdots,j_N$ represent the job tokens and $c_1,\cdots,c_M$ represent the course tokens. Then we add a MLP layer on top of the first output [CLS] embedding to obtain the relevance score of the job-course pair. 

\ipara{SuperGraphSAGE Model} aggregates the embeddings of the nodes' neighbors to represent them in a supervised end-to-end fashion. Specifically, given the job-word-course graph $G=(V,E)$, we invoke the bert-as-server API\footnote{https://github.com/hanxiao/bert-as-service} to generate the features for all nodes in $G$. For each job $j$, at the $l$-th convolutional time, it aggregates the embeddings of all its neighbors to obtain its new embedding $\mathbf{h}_{j}^{l}$, i.e., $\mathbf{h}_{j}^{l}$=$\sigma (\mathbf{W}^l \cdot {\rm CONCAT} (   \mathbf{h}_{j}^{l-1}  ,   \mathbf{h}^{l}_{\mathcal{N}(j)}  )  )$,  where $\sigma$ is a nonlinear function, $\mathbf{W}^l$ is the parameter matrix, $\mathbf{h}^{l}_{\mathcal{N}(j)} $ is the aggregated embedding of the job's neighbors, $\mathbf{h}_{j}^{l-1}$ is the previous embedding of the job, and CONCAT is the concatenate operation. We can obtain the $l$-th course embedding $\mathbf{h}_{c}^{l}$ in the same way. The cosine similarity between the final $L$-th embeddings of the course and the job can be viewed as their relevance score $r(j,c)$.

We use the loss function, $\mathcal{L} = \sum_{j,c+}   \log \sigma(r(j, c^+)) + \sum_{j,c-}  \log(1-\sigma(r(j, c^-)))$, to train the supervised models based on the pseudo labels provided by the unsupervised models. For a new $(j,c)$ pair in the test set, we estimate their relevance $r(j,c)$ as the average of all the relevance scores predicted by the $N^s$ supervised models, i.e., $r(j,c) =\frac{1}{N^s} \sum_{i=1}^{N^s} r_i(j,c)$.

\subsection{Automated Model Search}

\label{sec:auto}
To avoid human efforts to determine the suitable weak signals and the supervised models, in this paper, we propose to automatically search the optimal configuration of the weak supervision model. 
As the weak supervision model is a sequential process that first trains the unsupervised ranking models, then aggregates their outputs as the pseudo labels and finally trains the supervised ranking models based on the pseudo labels, we formalize the controller as a three-step LSTM model to sequentially determine which unsupervised models to select, which value of top-$k$ to select and which supervised models to select. The controller maintains a representation for each choice of different components, i.e., each unsupervised model, each value of $k$ and each supervised model. At step $t$, the representations of all the selections at $t-1$ are viewed as the input $\mathbf{x}_{t} \in \mathbb{R}^{d'}$,  which is taken together with the previous hidden vector $\mathbf{h}_{t-1} \in \mathbb{R}^{d'}$ and the cell state $\mathbf{e}_{t-1} \in \mathbb{R}^{d'}$ to produce the hidden vector $\mathbf{h}_{t}$  and the cell state $\mathbf{e}_{t}$, i.e.,

\beq{
\label{eq:lstm}
\mathbf{h}_t, \mathbf{e}_t = \mathbf{LSTM}(\mathbf{x}_t, \mathbf{h}_{t-1}, \mathbf{e}_{t-1}, \Phi),
}

\noindent where $\Phi$ are the parameters of LSTM and $d'$ is the hidden vector dimension. Finally, the component at step $t$ is determined according to  the hidden vector $\mathbf{h}_t$ and the representations of all the choices of the component at time $t$. Now we present the details of the sampling process:

\ipara{Step 1: Unsupervised model sampling} 
is to sample unsupervised models. At the beginning, the controller selects none of the components and has no memory, thus we set the initial hidden state $\mathbf{h}_0$, the cell state $\mathbf{e}_0$ as empty embeddings, and randomly initialize the input $\mathbf{x}_1$. The controller takes $\mathbf{h}_0$, $\mathbf{e}_0$ and $\mathbf{x}_1$ as input, and output $\mathbf{h}_1$ and $\mathbf{e}_1$ by Eq.~\eqref{eq:lstm}. Then given $\mathbf{h}_1$, the controller samples the unsupervised models. Intuitively, an unsupervised model is more likely to be sampled if it is more related to the hidden vector $\mathbf{h}_1$ at this step. We randomly initialize the representation $\mathbf{w}_i \in \mathbb{R}^{d'}$ for each unsupervised model, multiply $\mathbf{w}_i$ with $\mathbf{h}_1$ to represent their relevance, based on which we perform sampling:

\beq{
	I^1_i  \sim  \text{softmax} (f( \mathbf{h}_1^T \times \mathbf{w}_i)),
}

\noindent where $\mathbf{h}_1^T \times \mathbf{w}_i$ is a $d'$-dimensional element-wise product between the two embeddings and $f$ is a fully-connected layer that converts the product into a 2-dimensional vector. Softmax is used to convert the vector into a probability distribution, from which the indicator variable $I^1_i \in \{0,1\}$ is sampled to represent whether the $i$-th unsupervised model should be selected or not. Essentially, we sample each model from a binomial distribution. After this step, we can get the indicator vector $\mathbf{I}^1 = [I^1_1, \cdots, I^1_{N^u}]$ to indicate the selected unsupervised models. For example in Fig.~\ref{fig:overview}, $\mathbf{I}^1 = [1 0 0 1 0 0 0 0]$ indicates the controller selects BM25 and LINE as the unsupervised models.

\ipara{Step2: $k$ sampling} is to sample the value of $k$ for selecting the top-$k$ ranked positive instances in the pseudo labels. 
After sampling the unsupervised models, we multiply the indicator vector $\mathbf{I}^1 = [I^1_1, \cdots, I^1_{N^u}]$ with the model representations $[\mathbf{w}_1, \cdots, \mathbf{w}_{N^u}]$ as the input $\mathbf{x}_2$ of the second step: 

\beq{
	\mathbf{x}_2 =  [I^1_1, \cdots, I^1_{N^u}] \cdot 	[\mathbf{w}_1, \cdots, \mathbf{w}_{N^u}]^T,
} 

\noindent where $\mathbf{x}_2$ denotes the summation of the representations of all the sampled models. With $\mathbf{x}_2$, $\mathbf{h}_1$ and $\mathbf{e}_1$ as the input, we can obtain $\mathbf{h}_2$ and $\mathbf{e}_2$ by Eq.~\eqref{eq:lstm}. 
Since the value space of $k$ can be very large, to simplify the sampling process, we first categorize all the values of $k$ into $\tau$ categories and sample one category for $k$. The sampling process is defined as:

\beq{
	\mathbf{I}^2 \sim  \text{softmax} (g( \mathbf{h}_2)),
}

\noindent where $g$ is a full-connected layer that converts the hidden vector $\mathbf{h}_2$ into a $\tau$-dimensional vector. Softmax is used to convert the vector into a probability distribution, from which the indicator vector $\mathbf{I}^2 \in \{0,1\}^{\tau}$ is sampled to represent which category of $k$ is selected. Note that $\mathbf{I}^2$ is a one-hot vector with only one dimension as one, whose index indicates the sampled category of $k$. Essentially, we sample the category of $k$ from a multinomial distribution. For example in Fig.~\ref{fig:overview}, $\mathbf{I}^2 = [10000]$ indicates the controller selects the first category for $k$ and its corresponding value is 20.

\ipara{Step 3: Supervised model sampling} is to sample supervised models. After sampling $k$, we multiply the sampling indicator $\mathbf{I}^2$ with the concatenation of the $k$'s category representations $[\mathbf{z}_1, \cdots, \mathbf{z}_{\tau}]$ as the input $\mathbf{x}_3$ of the third step: 

\beq{
	\mathbf{x}_3 =  [I_1^2, \cdots, I_{\tau}^2] \cdot 	[\mathbf{z}_1, \cdots, \mathbf{z}_{\tau}]^T,
} 

\noindent where the category representations $[\mathbf{z}_1, \cdots, \mathbf{z}_{\tau}]$ for each category of $k$ are randomly initialized. The input $\mathbf{x}_3$ denotes the representation of the selected category.  With $\mathbf{x}_3$, $\mathbf{h}_2$ and $\mathbf{e}_2$ as the input, we can obtain $\mathbf{h}_3$ and $\mathbf{e}_3$ by Eq.~\eqref{eq:lstm}. Given $\mathbf{h}_3$, we can sample the indicator $\mathbf{I}^3$ following the same sampling process of step 1 to determine which supervised models should be selected, i.e., $	I^3_i  \sim  \text{softmax} (q( \mathbf{h}_3^T \times \mathbf{u}_i))$, where $q$ is a fully-connected layer that converts the product $\mathbf{h}_3^T \times \mathbf{u}_i$ into a 2-dimensional vector, $\mathbf{u}_i \in \mathbb{R}^{d'} $ is a randomly initialized embedding for the $i$-th supervised model. As shown in Fig.~\ref{fig:overview}, $\mathbf{I}^3=[00010]$ indicates the controller selects BERT interaction model.

\subsection{Reinforcement Joint Training}
Once the controller finishes searching the configurations of the weak supervision model, i.e.,  the unsupervised models, the top-$k$ value and the supervised models, a combination with this architecture is built and trained. When the searching architecture achieves convergence, it will get an accuracy $\mathcal{R}$ on a small hold-out annotated dataset (validation set). The accuracy $\mathcal{R}$ is viewed as reward and the parameters of the controller LSTM are then optimized  in order to search the best configurations that can achieve the maximal expect validation accuracy. In this paper, we propose a reinforcement joint training process to update the parameters of the controller LSTM, denoted by $\Phi$, and the parameters of the weak supervision model, denoted by $\Theta$. The reinforcement joint training process consists of two interleaving phrases (Algorithm~\ref{algo:rl}), the first phrase trains $\Theta$, while the second phrase trains $\Phi$,  the details are as follows:

\vpara{Training $\Theta$.}
When training $\Theta$, we fix the controller's sampling policy $\pi(\mathbf{m}; \Phi)$, i.e., the three-step sampling strategy, and perform stochastic gradient descent on $\Theta$ to maximize the expected loss $\mathbb{E}_{\mathbf{m} \sim \pi}[ \mathcal{L}(\mathbf{m};\Theta)]$, where $\mathcal{L}(\mathbf{m};\Theta)$ is the loss computed on a minibatch of training data, with a weak supervision model $\mathbf{m}$ sampled from $\pi(\mathbf{m}; \Phi)$. The gradient is computed using Monte Carlo estimate:

\beq{
\label{eq:train_theta}
\nabla_\Theta  \mathbb{E}_{\mathbf{m} \sim  \pi}[\mathcal{L}(\mathbf{m};\Theta)] \!\approx \!\frac{1}{N_m} \sum_{p=1}^{N_m}\! \sum_{q=1}^{N_u'+N_s'}\!\nabla_\Theta  \mathcal{L}_q(\mathbf{m}_p, \Theta),
}

\normalem
\begin{algorithm}[t]
	{\small \caption{Reinforcement Jointly Training\label{algo:rl}}
		\KwIn{ A set of jobs $J$ and a set of courses $C$.}
		\KwOut{Parameters $\Theta$ of the weak supervision model and $\Phi$ of the controller.}
		Initialize $\Phi = \Phi^0$, $\Theta = \Theta^0$;\\ 
		Pre-train the $N^u$ unsupervised ranking models;\\
		\Repeat{Convergence}{
			Sample a weak supervision model $\mathbf{m}$ from $\pi(\mathbf{m};\Phi)$; \\
			Train $\Theta$ of $\mathbf{m}$ by Eq.~\eqref{eq:train_theta};\\
			Calculate $\mathcal{R}^s+\mathcal{R}^u$ of $\mathbf{m}$ on the validation set;\\
			Update $\Phi$ in the controller by REINFORCE;\\
		}
		
}\end{algorithm}
\ULforem

\noindent where $N_m$ is the sampling times of the weak supervision model in one epoch. It is empirically proven that $N_m$=1 works just fine~\cite{pham2018efficient}. Notations $N_u'$ and $N_s'$ represent the number of the sampled unsupervised and the supervised models respectively. The whole loss $\mathcal{L} (\mathbf{m}_p, \Theta)$ is the summed losses of all the sampled models. No matter which unsupervised ranking models are sampled, they are always trained on the same training data. So we can pre-train each unsupervised ranking model on the training data, and directly fetch the relevance scores between jobs and courses from all the sampled unsupervised models during each update of Eq.~\eqref{eq:train_theta}. Thus after each sampling of $\bf{m}$, we only need to re-optimize the loss $\mathcal{L}_q$ of each sampled supervised model. As a result, the loss $\mathcal{L} (\mathbf{m}_p, \Theta)$ is the summed losses of all the sampled supervised models.

\vpara{Training $\Phi$.}
When training $\Phi$, we fix $\Theta$ of the weak supervision model and perform REINFORCE algorithm~\cite{williams1992simple} on $\Phi$ to maximize the expected reward $\mathbb{E}_{\mathbf{m} \sim \pi(\mathbf{m};\Phi)}[\mathcal{R}(\mathbf{m}, \Theta)]$. 
The actions of the controller are to sample the unsupervised models, $k$, and the supervised models sequentially. The reward is regarded as the evaluated mean reciprocal rank (MRR) of  the sampled supervised models on the validation set, which is the set of a few job-course pairs annotated by human beings. Besides, the MRR achieved by the aggregation results of the sampled unsupervised models can also be regarded as the additional reward to accelerate the training process~\cite{ghavamzadeh2003hierarchical}.  Thus, the final reward is defined as $\mathcal{R} = \mathcal{R}^s+\mathcal{R}^u$, where $\mathcal{R}^s$ and $\mathcal{R}^u$ are the rewards from the supervised and unsupervised models.

\section{Experiment}
\label{exp}
In this section, we evaluate our proposed model \textit{AutoWeakS} against several unsupervised, supervised, and weak supervision baselines. We also explore whether the selections for each component in \textit{AutoWeakS}  (i.e., the selections for the unsupervised models, the supervised models and the top-$k$ values) are necessary.

\subsection{Experimental Setup} 

\begin{table*}
	\newcolumntype{?}{!{\vrule width 1pt}}
	\newcolumntype{C}{>{\centering\arraybackslash}p{4em}}
	\caption{
		\label{tb:performance} Overall performance of recommending courses for jobs. {\small We try different $k$ for WeakS and report its best performance. }
		\normalsize
	}
	\centering  \small
	\renewcommand\arraystretch{1.0}
	\begin{tabular}{@{~}l@{~}?*{1}{CCC?}*{1}{CCC}}
		\toprule
		\multirow{2}{*}{\vspace{-0.3cm} Model}
		&\multicolumn{3}{c?}{JD-XuetangX}
		&\multicolumn{3}{c}{Tencent-XuetangX} 
		
		\\
		\cmidrule{2-4} \cmidrule{5-7} 
		& {HR@5} &  {NDCG@5} &  {MRR} & {HR@5} & {NDCG@5} & {MRR}\\
		\midrule
		BM25 
		& 0.162  & 0.151   & 0.173	
		& 0.072 & 0.046   & 0.070	
		\\
		Word2vec 
		&0.301&0.212&0.217
		&0.142&0.107&0.114
		\\
		BERT
		&0.348&0.239&0.238
		&0.159&0.104&0.122
		\\
		LINE 
		& 0.489  & 0.362  & 0.409	
		& 0.396 & 0.284   & 0.279	
		\\  
		PTE 
		&0.378    & 0.244    & 0.334	
		&0.295  & 0.204   & 0.210	
		\\
		DeepWalk
		& 0.390  & 0.249  & 0.258	
		& 0.370  & 0.262   & 0.261	
		\\
		Node2vec
		& 0.374  & 0.279   & 0.284	
		& 0.386  & 0.282  & 0.277	
		\\
		GraphSAGE
		&0.312&0.252&0.232
		&0.186&0.121&0.139
		\\
		\midrule
		Traditional Representation
		&0.407&0.261&0.262
		&0.201&0.125&0.148
		\\
		Traditional Interaction
		&0.470&0.429&0.414
		&0.324&0.215&0.214
		\\
		BERT Representation
		&0.350&0.232&0.231
		&0.294&0.195&0.204
		\\	
		BERT Interaction
		&0.564&0.537&0.497
		&0.405&0.254&0.222
		\\
		SuperGraphSAGE
		&0.263&0.176&0.186
		&0.231&0.144&0.155
		\\
		\midrule
		WeakS
		& 0.704  & 0.548   & 0.592 
		& 0.370 & 0.255   & 0.227	
		\\
		LINE+AllS
		&0.736&0.550&0.624
		&0.408&0.289&0.236
		\\
		\midrule
		\textbf{AutoWeakS}
		&\textbf{0.793}&\textbf{0.615}&\textbf{0.671}
		&\textbf{0.631}&\textbf{0.522}&\textbf{0.540}
		\\
		
		\bottomrule
		
	\end{tabular}
	
\end{table*}

\vpara{Dataset.}
We collect all the courses from XuetangX\footnote{http://www.xuetangx.com}, one of the largest MOOCs in China, and this results in 1951 courses. The collected courses involve seven areas: computer science, economics, engineering, foreign language, math, physics, and social science. Each course contains 131 words in its descriptions on average. We also collect 706 job postings from the recruiting website operated by JD.com\footnote{http://campus.jd.com/home} (JD) and 2,456 job postings from the website owned by  Tencent corporation\footnote{https://hr.tencent.com/} (Tencent). The collected job postings involve six areas: technical post, financial post, product post, design post, market post, supply chain and engineering post. Each job contains 107 and 151 words in its posting on average in JD and Tencent respectively. To evaluate the model performance, for both JD and Tencent dataset, we randomly select 200 jobs, and ask ten volunteers to annotate the relevant courses to the jobs. Specifically, for a queried job, we first use each unsupervised model  in Section~\ref{sub:unsupervised-model} to calculate a relevance score for each course, average all the scores over all the models, select top 60 candidate courses, annotate each candidate and obtain the ground truth by majority voting of all the volunteers' annotations. The Dataset and the code are online now\footnote{https://github.com/jerryhao66/AutoWeakS}.

\hide{\footnote{https://github.com/anonymous-x1/AutoWeakS}.}

\vpara{Settings.}
For training the unsupervised ranking models, we use all the 706 job postings from JD and all the 2,456 job postings from Tencent to learn the embeddings of the jobs and courses.
For the supervised ranking models, we hold out the human annotated jobs and only use the 506 unlabeled jobs from JD and 2,256 unlabeled jobs from Tencent for training. 
On each dataset, we averagely partition the annotated 200 jobs into a validation set and a test set and sample 99 negative instances for each positive instance (1 positive plus 99 negatives)~\cite{he2018nais}. We use Hit Ratio of top $K$ items (HR@$K$), Normalized Discounted Cumulative Gain of top $K$ items (NDCG@$K$) and Mean Reciprocal Rank (MRR) as the evaluation metrics for ranking, where $K$ is set as 5. 

\subsection{Experimental Results}
\vpara{Comparison with Baselines.}
In this experiment, we evaluate our model \textit{AutoWeakS} against the unsupervised and supervised models in Section~\ref{sub:unsupervised-model}, the weak supervision model WeakS, which includes all the unsupervised models and the supervised models without model search, and one competitive baseline, LINE+AllS, which includes LINE as the unsupervised model and all the supervised models. Note that due to the lack of enough labeled data,  we only use a small  annotation data  (i.e., the validation set) to train the supervised models.

Table \ref{tb:performance} shows the results on two datasets. We vary the value $k$ for WeakS, LINE+AllS and report its best performance in Table \ref{tb:performance}. From the results, we can see that the proposed \textit{AutoWeakS} performs clearly better than other baselines. Compared with the unsupervised graph-based matching methods, unsupervised text-only matching models perform worse, as only using the descriptive words of the jobs and the courses can not capture high-order relationships between the jobs and the courses. Some supervised methods such as BERT interaction model and traditional interaction model perform better than the unsupervised methods, as the unsupervised methods do not explicitly compare the relevance of the positive courses and the negative courses to a queried job. However, due to the lack of enough training labels, the performance of the supervised models is worse than WeakS, LINE+AllS and our proposed method \textit{AutoWeakS}.

\begin{table*}
	\newcolumntype{?}{!{\vrule width 1pt}}
	\newcolumntype{C}{>{\centering\arraybackslash}p{4em}}
	\caption{
		\label{tb:compare_unsupervised} Performance of different choices of unsupervised models in AutoWeakS with $k$ and the supervised component fixed. 
		\normalsize
	}
	\centering  \small
	\renewcommand\arraystretch{1.0}
	\begin{tabular}{@{~}l@{~}?*{1}{@{}CCC?}*{1}{CCC}@{}}
		\toprule
		\multirow{2}{*}{\vspace{-0.3cm} Unsuper. choices}
		&\multicolumn{3}{c?}{JD-XuetangX}
		&\multicolumn{3}{c}{Tencent-XuetangX} 
		
		\\
		\cmidrule{2-4} \cmidrule{5-7} 
		& {HR@5} &  {NDCG@5} & {MRR} & {HR@5} & {NDCG@5}  & {MRR}\\

		\midrule
		BM25+
		& 0.203  & 0.194  & 0.182	
		&0.183&0.139&0.159
		\\	
		Word2vec+
		&0.435&0.392&0.336
		&0.333&0.312&0.321
		\\
		BERT+
		&0.705&0.511&0.511
		&0.393&0.373&0.387
		\\
		LINE+
		& 0.722 & 0.559 & 0.516
		& 0.589   & 0.478    &0.499		
		\\ 
		PTE+
		&0.657   & 0.488    & 0.505
		& 0.471  & 0.451   &0.497 	
		\\
		DeepWalk+
		& 0.677  & 0.503 & 0.462	
		& 0.508  & 0.461  & 0.411
		\\
		Node2vec+
		& 0.684  & 0.507  & 0.451	
		& 0.534 & 0.449 &0.463 
		\\
		GraphSAGE+
		&0.642&0.495&0.402
		&0.563&0.471&0.491
		\\
		\midrule
		All unsupervised+
		&0.609&0.423&0.458
		&0.415&0.396&0.426
		\\
		
		\midrule
		\textbf{AutoWeakS}
		&\textbf{0.793}&\textbf{0.615}&\textbf{0.671}
		&\textbf{0.631}&\textbf{0.522}&\textbf{0.540}
		\\
		\bottomrule
	\end{tabular}
	
\end{table*}

WeakS performs better than all the unsupervised models on JD-XuetangX, as it explicitly learns the ranking of the candidate courses to queried jobs. However, on Tencent-XuetangX, WeakS underperforms several unsupervised models, because BM25 performs particularly poorly on this dataset, which reduces the effect of the aggregated pseudo labels from all the unsupervised models. This also indicates that indiscriminately combing all the models may not result in the best performance. Besides, \textit{AutoWeakS} beats LINE+AllS, which implies selecting only one unsupervised model may suffer from the issue of ranking bias. 

We further remove BM25 from WeakS, name the model as WeakS-BM25 and show the performance of WeakS, LINE+AllS and WeakS-BM25 in Fig.~\ref{subfig:small_data_hr5} and Fig.~\ref{subfig:big_data_hr5}. The $k$ value of \textit{AutoWeakS} is fixed as the automatically searched value. The results show that even if the worst performed BM25 is removed,  given any value of $k$, WeakS-BM25 still underperforms \textit{AutoWeakS}, which indicates the advantage of the automated model search in \textit{AutoWeakS}.

\hide{ For training \textit{AutoWeakS}, at each epoch, it takes only a few seconds to train the controller, about 30 minutes to train the sampled weak supervision model on JD-XuetangX and 1 hours on Tencent-XuetangX. The controller can converge after 30 epochs. }

\begin{figure}[t]
	\centering
	\mbox{ 
		\subfigure[\scriptsize JD-XuetangX]{\label{subfig:small_data_hr5.pdf}
			\includegraphics[width=0.22\textwidth]{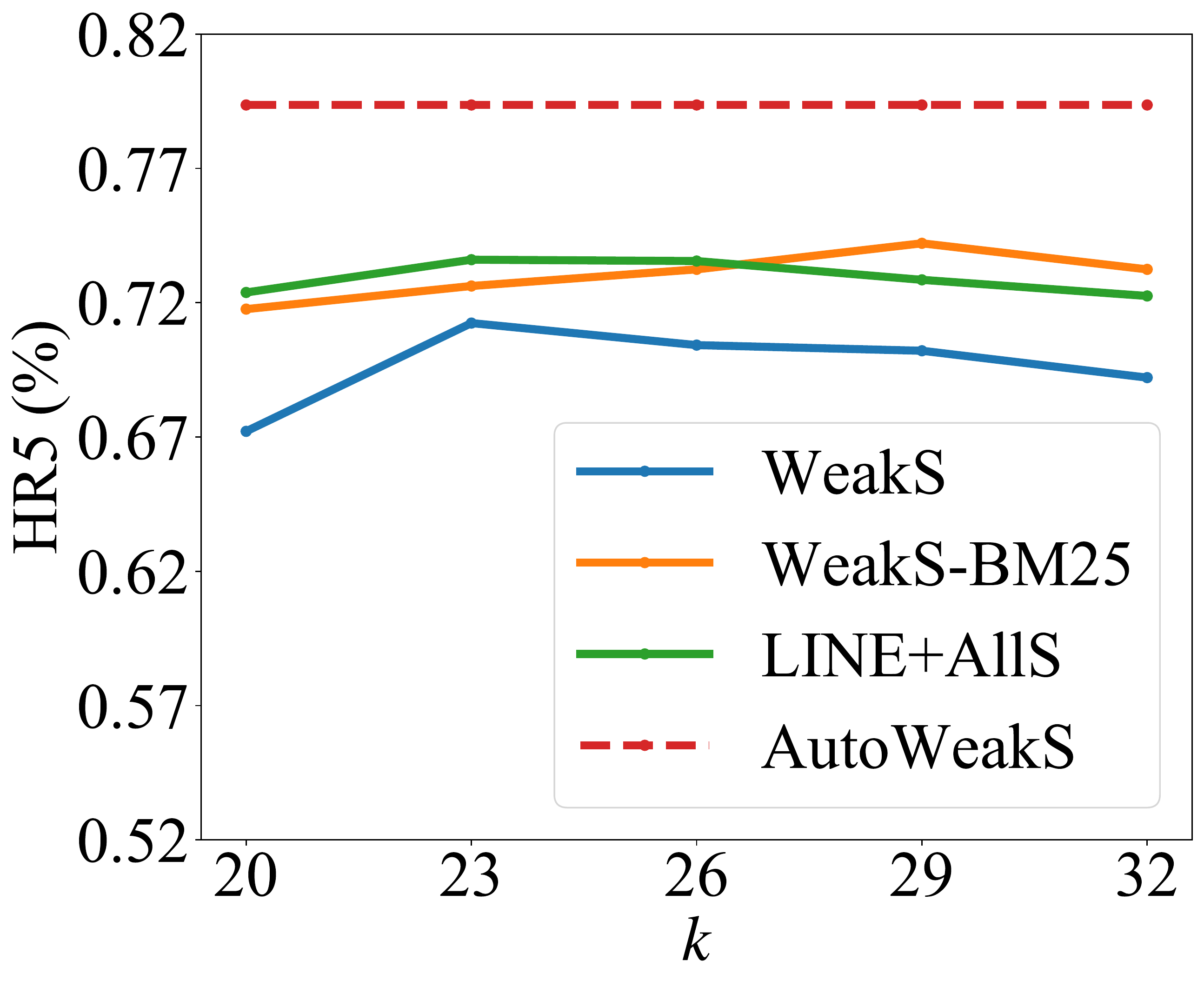}
		}
		
		\subfigure[\scriptsize Tencent-XuetangX]{\label{subfig:big_data_hr5.pdf}
			\includegraphics[width=0.22\textwidth]{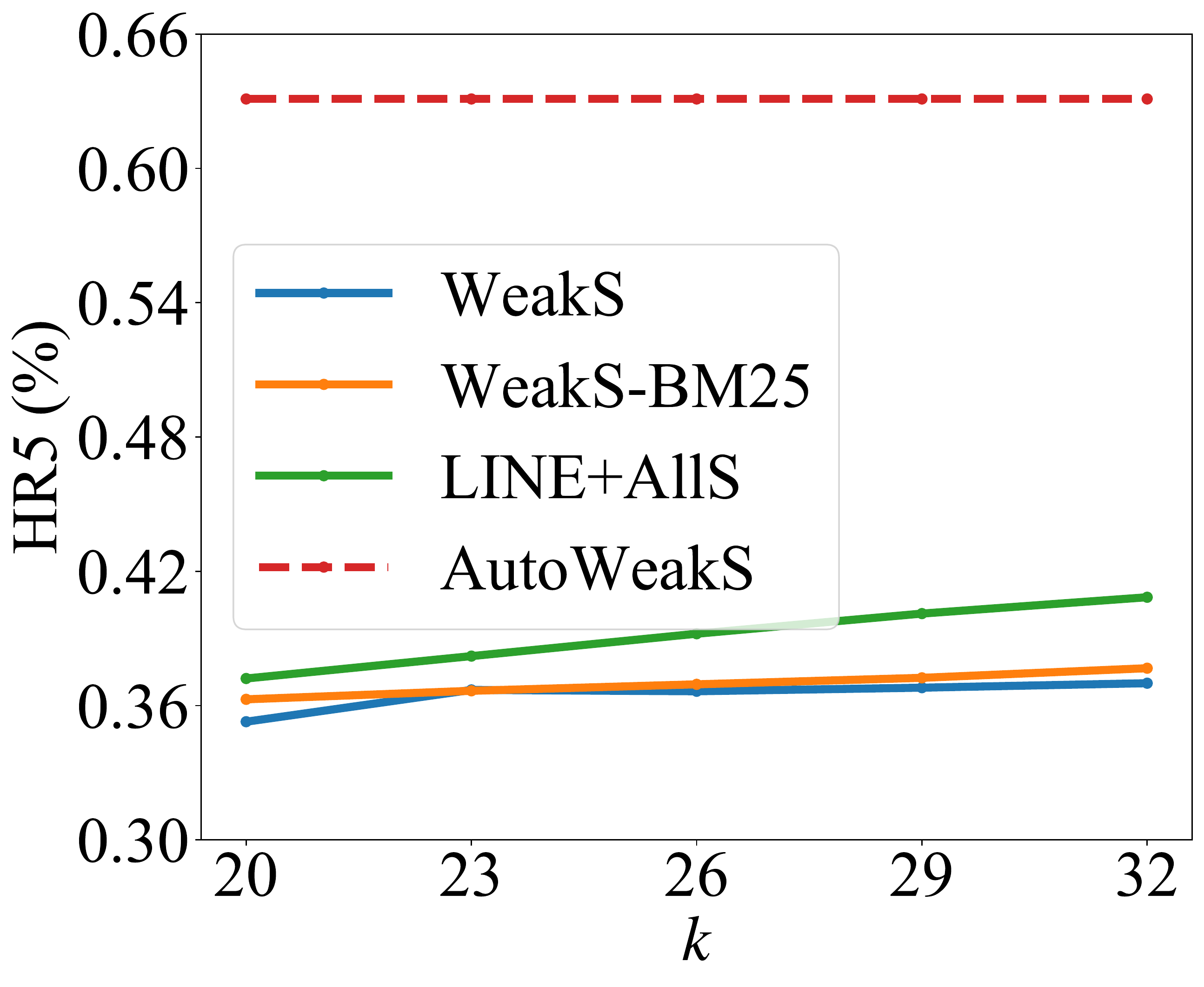}
		}
		
	\subfigure[\scriptsize JD-XuetangX]{\label{subfig:jd_k.pdf}
		\includegraphics[width=0.22\textwidth]{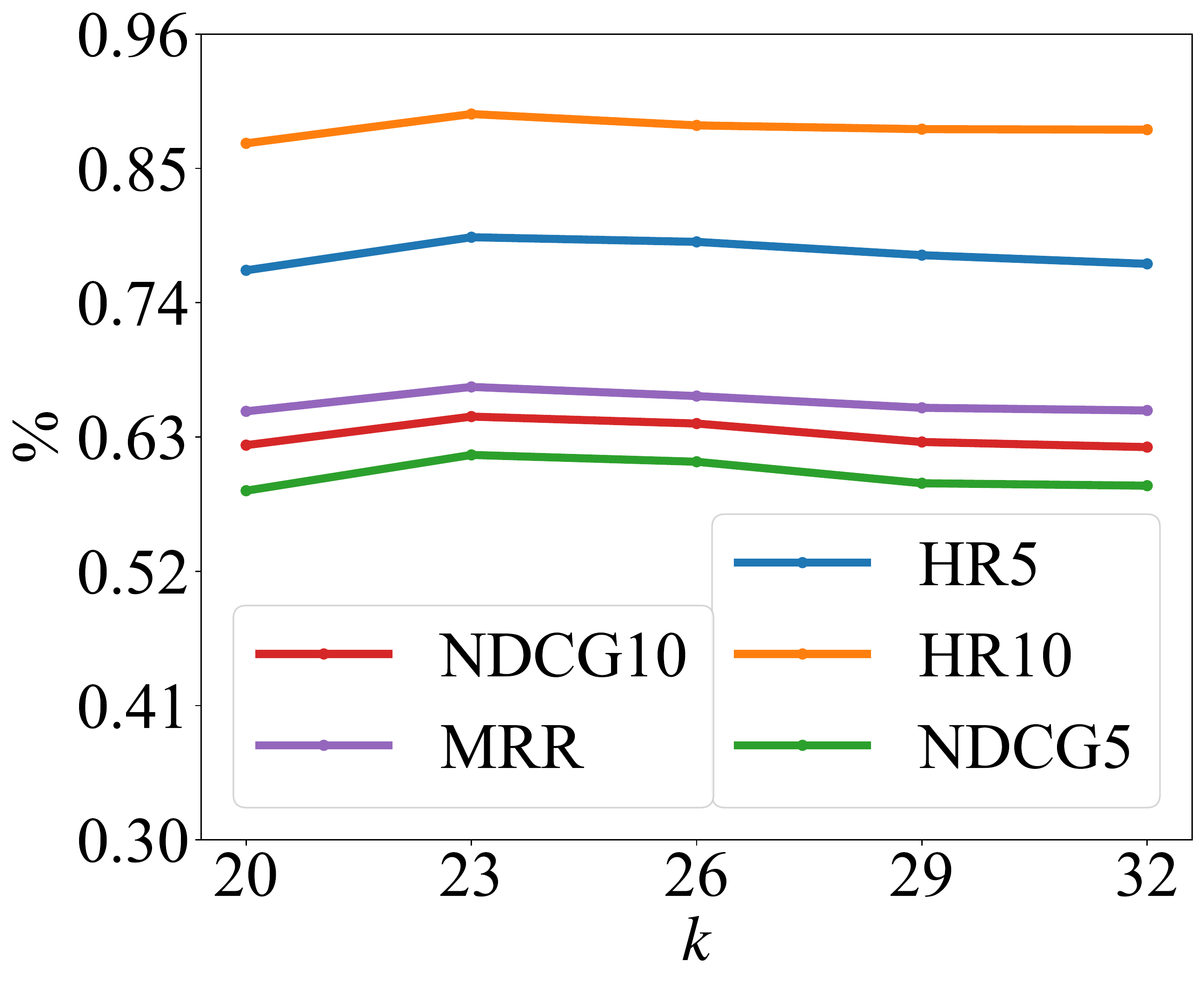}
	}
	
	\subfigure[\scriptsize Tencent-XuetangX]{\label{subfig:tx_k.pdf}
		\includegraphics[width=0.22\textwidth]{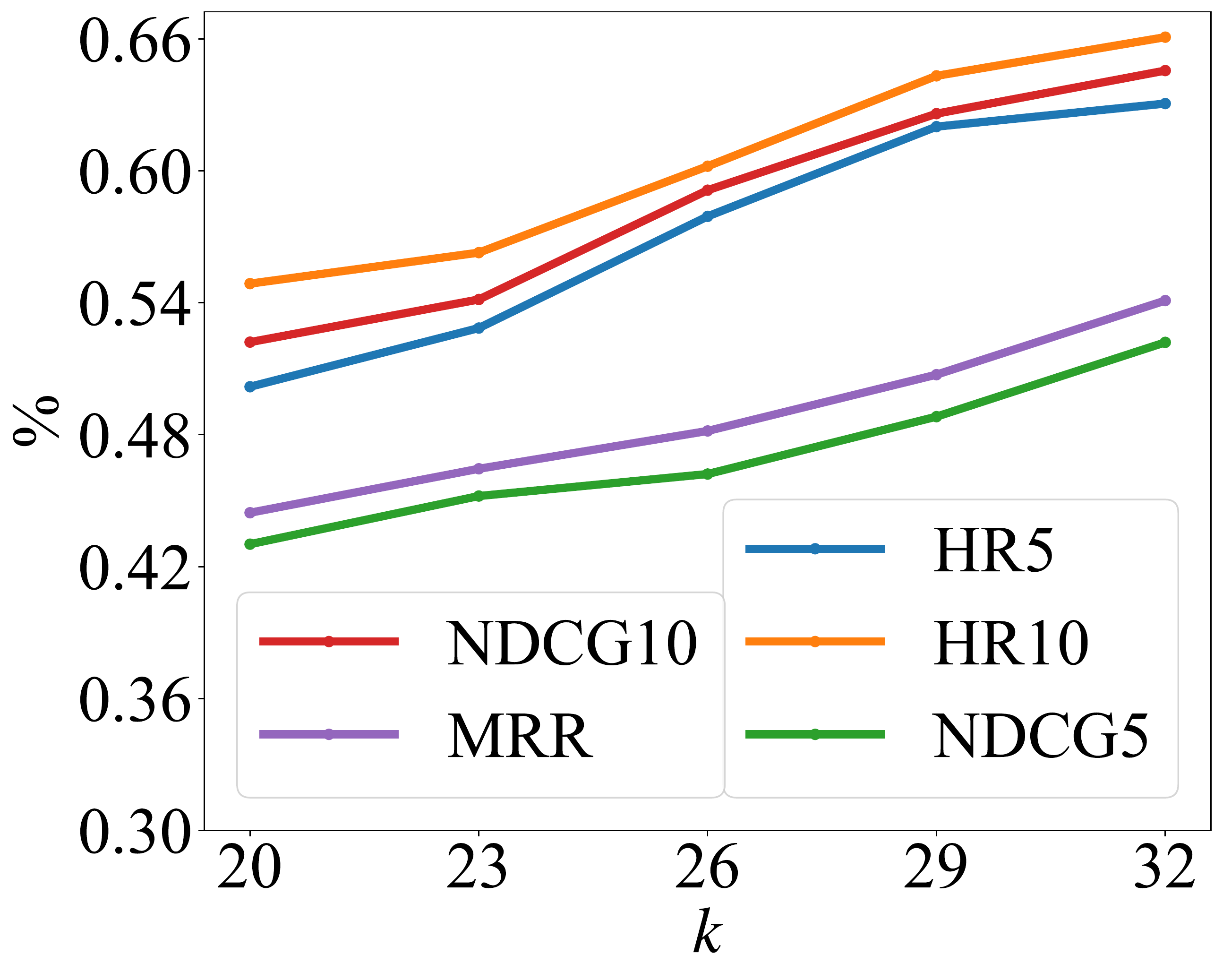}
	}
	}
	
	\caption{\label{fig:arc_jd} (a), (b) show that the baselines with any choice of top-$k$ underperform AutoWeakS with the automatically searched top-$k$ value. (c), (d) further present the results of AutoWeakS under different top-$k$ values, which indicates that the automatically searched top-$k$ value performs the best against all the other top-$k$ values. }
\end{figure}

\begin{figure}[t]
	\centering
	\mbox{ 
		\subfigure[\scriptsize JD-XuetangX]{\label{subfig:small_arc.pdf}
			\includegraphics[width=0.28\textwidth]{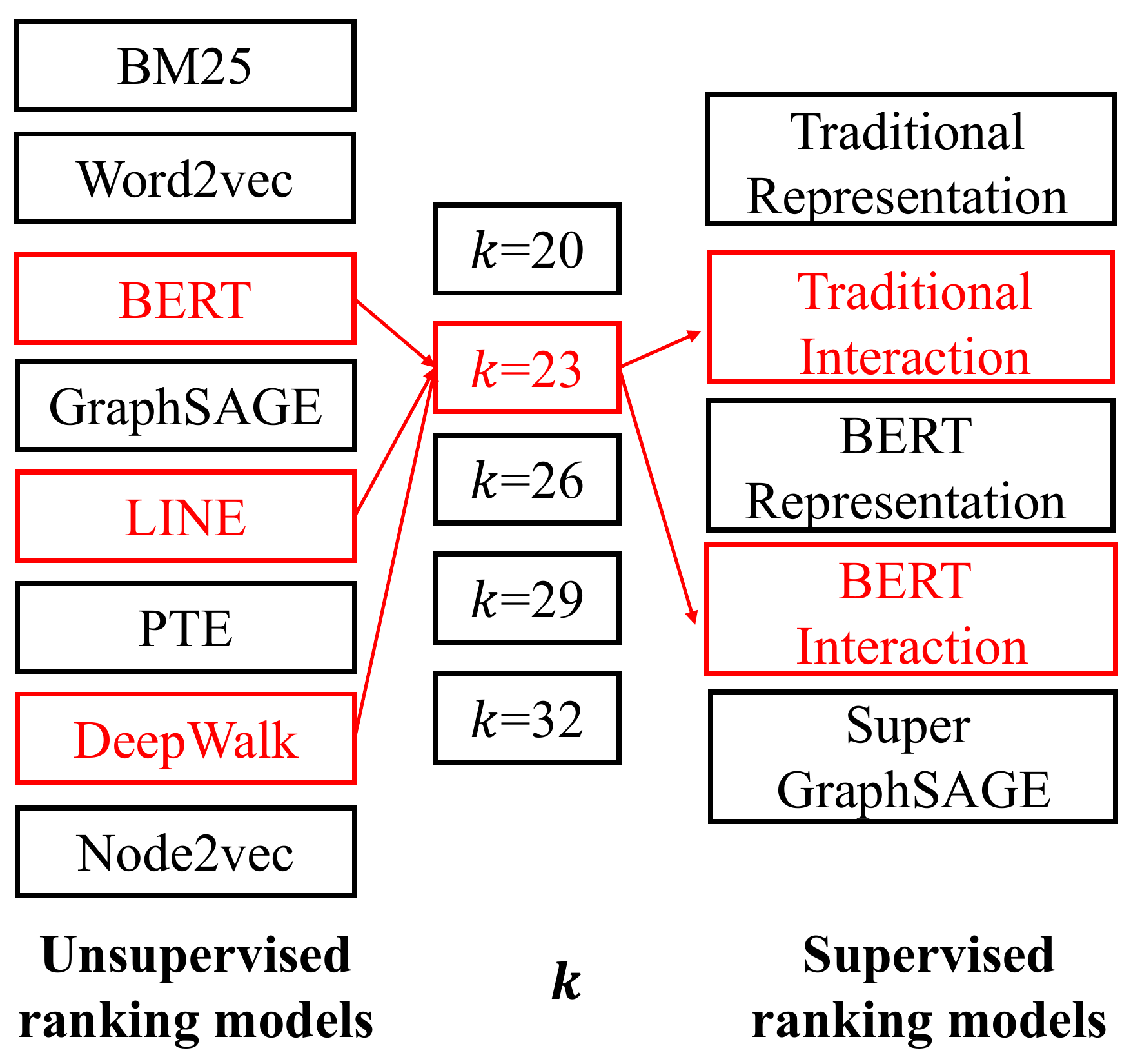}
		}
		
		\hspace{0.25in}
		\subfigure[\scriptsize Tencent-XuetangX]{\label{subfig:big_arc.pdf}
			\includegraphics[width=0.28\textwidth]{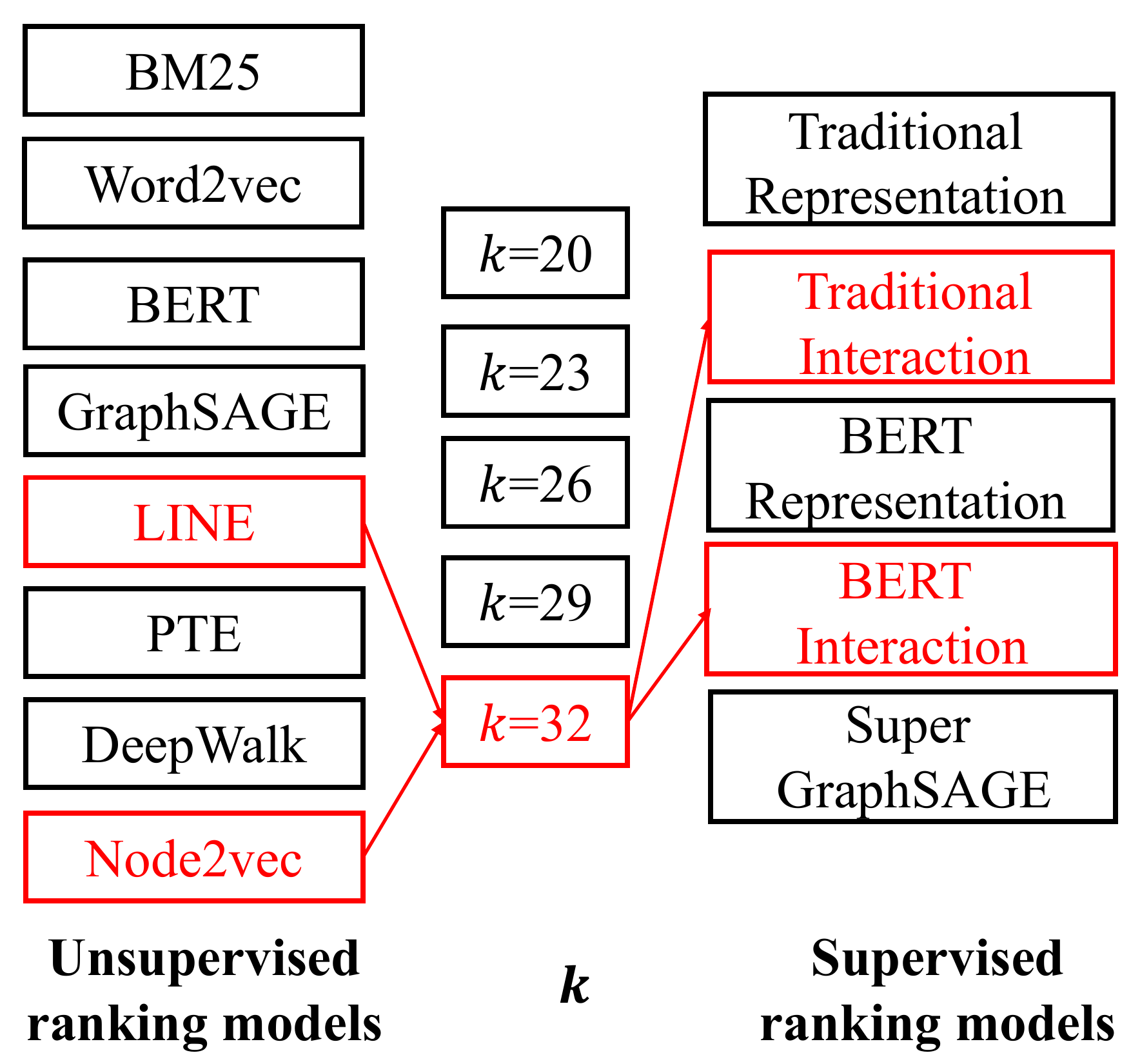}
		}
	
	}
	
	\caption{\label{fig:arc} The optimal selections of AutoWeakS.}
\end{figure}

\begin{table*}
	\newcolumntype{?}{!{\vrule width 1pt}}
	\newcolumntype{C}{>{\centering\arraybackslash}p{4em}}
	\caption{
		\label{tb:compare_supervised_autos} Performance of different choices of supervised models in AutoWeakS with the unsupervised component and $k$ fixed. 
		\normalsize
	}
	\centering  \small
	\renewcommand\arraystretch{1.0}
	\begin{tabular}{@{~}l@{~}?*{1}{CCC?}*{1}{CCC}}
		\toprule
		\multirow{2}{*}{\vspace{-0.3cm} Super. choices}
		&\multicolumn{3}{c?}{JD-XuetangX}
		&\multicolumn{3}{c}{Tencent-XuetangX} 
		
		\\
		\cmidrule{2-4} \cmidrule{5-7} 
		& {HR@5} & {NDCG@5} & {MRR} & {HR@5} & {NDCG@5} & {MRR}\\
		\midrule 
		Traditional Representation+
		& 0.525   & 0.349   & 0.335	
		& 0.362 &0.283  & 0.269	
		\\ 
		Traditional Interaction+
		& 0.679 &0.508  &0.483 
		& 0.601 &0.502& 0.504 
		\\
		BERT Representation+
		&0.604&0.483&0.462
		&0.318&0.209&0.209
		\\
		BERT Interaction+
		&0.729&0.537&0.609
		&0.576&0.464&0.481
		\\
		SuperGraphSAGE+
		&0.348&0.212&0.368
		&0.265&0.181&0.193
		\\
		\midrule
		All supervised+
		&0.652&0.482&0.507
		&0.402&0.317&0.301
		\\
		\midrule
		\textbf{AutoWeakS}
		&\textbf{0.793}&\textbf{0.615}&\textbf{0.671}
		&\textbf{0.631}&\textbf{0.522}&\textbf{0.540}
		\\
		\bottomrule
	\end{tabular}
	
\end{table*}

\vpara{Analysis of Unsupervised Component.}
We evaluate the performance of different choices of the unsupervised models, when fixing the sampled $k$ and the supervised component in \textit{AutoWeakS}. We name the model as BM25+ if only BM25 is chosen to produce pseudo labels. Other single model is named in the same way. All unsupervised+ means we combine the labels of all the unsupervised models. 
Fig.~\ref{subfig:small_arc} shows that on JD-XuetangX, \textit{AutoWeakS} selects the combination of BERT, LINE, and DeepWalk, which performs better than any single unsupervised model and All unsupervised+ shown in Table~\ref{tb:compare_unsupervised}. 
On Tencent-XuetangX, \textit{AutoWeakS} also obtains the best performance, and it selects the combination of LINE and Node2vec as the unsupervised component.
The results indicate the advantage of automatically searching the unsupervised models.

\vpara{Analysis of $k$.}
We evaluate the performance of different choices of $k$ to generate the pseudo labels, when fixing the sampled unsupervised and the supervised components in \textit{AutoWeakS}. Fig.~\ref{fig:arc} presents that the automatically searched $k$ is 23 on JD-XuetangX and is 32 on Tencent-XuetangX. Fig.~\ref{subfig:jd_k} and Fig.~\ref{subfig:tx_k} show that \textit{AutoWeakS} with other $k$ values underperforms the searched $k$ values. The results indicate the advantage of automatically searching $k$.

\vpara{Analysis of Supervised Component.}
We evaluate the performance of different choices of the supervised models, when fixing the sampled unsupervised component and $k$ in \textit{AutoWeakS}. We name the model as BERT Interaction+ if only the BERT interaction model is trained. Other single supervised model is named in the same way. All supervised+ means we train all the supervised models.
Fig.~\ref{subfig:small_arc} and Fig.~\ref{subfig:big_arc} show that on both of the JD-XuetangX and the Tencent-XuetangX datasets, \textit{AutoWeakS} selects the combination of the BERT interaction model and the traditional interaction model. The results show that \textit{AutoWeakS} performs better than all the other choices shown in Table~\ref{tb:compare_supervised_autos}, which indicates the advantage of automatically searching the supervised models.


\section{Related Work}
\label{related}

Much effort has been made to provide better services for job seekers and recruiters through analyzing the flow of job seekers~\cite{Oentaryo2018} or matching the job recruitment postings and the resumes of the job seekers~\cite{Tong2017Measuring}. The related works include:

\vpara{Weak Supervision Model.}
Training neural ranking models on pseudo-labeled data has been attracted attentions. For example, Dehghani et al.~\cite{dehghani2017neural} leverage the output of traditional IR models such as BM25 as the weak supervision signal to generate a large amount of pseudo labels to train effective neural ranking models. Zamani et al.~\cite{zamani2018neural} train a neural query performance predictor by multiple weak supervision signals, and they also provide a theoretical analysis of this weak supervision method~\cite{zamani2018theory}. The same idea is employed in~\cite{dehghani2017avoiding,luo2017training}. However, for different tasks, human efforts are demanded to determine the suitable weak signals and the supervised models. Even if each signal is carefully selected by humans, their combination may not be optimal. 

\vpara{Automated Machine Learning (AutoML).}
The goal of AutoML is to automatically determine the optimal configurations such as selecting the optimal models~\cite{feurer2015efficient,kotthoff2017auto}, features~\cite{katz2016explorekit,huang2019efficient}, and neural architecture~\cite{zoph2016neural,liu2018progressive}, which can help people use machine learning models easily. Different types of techniques are studied to search the optimal configuration. For example, Bayesian optimization methods such as Auto-sklearn~\cite{feurer2015efficient} and Auto-Weka~\cite{kotthoff2017auto} model the relationship between a configuration and the corresponding performance in a probabilistic way. Reinforcement learning trains the optimal search policies according to the feedbacks of the searched configurations~\cite{sutton2018reinforcement}, where the search policy can be modeled by RNN~\cite{pham2018efficient,zoph2016neural}.  Inspired by the above works, we propose a RL-based joint training framework to search an optimal combination of the unsupervised/supervised models and the hyperparameter $k$ in the proposed weak supervision model for recommending courses for jobs.

\section{Conclusion}
\label{conclusion}
We present the first attempt to solve the problem of recommending courses in MOOCs for jobs \hide{in online recruitment platform}by a general automated weak supervision model. With reinforcement joint training of a weak supervision model for recommending courses and a controller for searching models, we can automatically find the best configuration of the weak supervision model. Experiments on two real-world datasets of jobs and courses show that the proposed AutoWeakS significantly outperforms the classical unsupervised, supervised and weak supervision baselines.

\section*{ACKNOWLEDGMENTS}
This work is supported by National Key R\&D Program of China (No.2018YFB1004401) and NSFC  (No.61532021, 61772537, 61772536, 61702522).

\end{document}